\documentstyle[12pt,epsf]{article}

\newcommand{\bq}{\begin{equation}}
\newcommand{\ee}{\end{equation}}

\newcommand{\fr}[2]{\frac{#1}{#2}}

\begin{document}
\pagestyle{plain}
\pagenumbering{arabic}

\parskip = 2ex

\vspace{0.6cm}

\begin{center}

{\Large \bf High order corrections to the Renormalon}\\

\vspace{0.4cm}

{\bf S. V. Faleev$^1$,} {\bf 
{P. G. Silvestrov}$^{1,2}$} \\
$^1${\it Budker Institute of Nuclear Physics, 630090 Novosibirsk,
Russia}\\ $^2${\it The Niels Bohr Institute, Blegdamsvej 17, DK-2100
Copenhagen {\O}, Denmark}

\vspace{0.5cm}

\end{center}
\begin{abstract}

The high order corrections to renormalon are considered. Each new type
of insertions into the renormalon chain of graphs generates the
correction to the asymptotics of perturbation theory of the order of
$\sim 1$. However, this series of corrections to the asymptotics is
not the asymptotic one (i.e. the $m$-th correction does not grow like
$m!$). The summation of these corrections for UV renormalon may change
the asymptotics by factor $N^\delta$. For the traditional IR
renormalon the $m$-th correction diverges like $(-2)^m$. However, this
divergency has no infrared origin and may be removed by proper
redefinition of IR renormalon. On the other hand for IR renormalons in
hadronic event shapes one should naturally expect these multi-loop
contributions to decrease like $(-2)^{-m}$. Some problems
expected upon reaching the best accuracy of perturbative QCD are also
discussed.

\end{abstract}

\section{Introduction}\label{sec1}

The running coupling constant seems to be an inevitable companion of
any nontrivial renormalizable field theory. In its turn the
asymptotics of perturbation theory for any quantity calculated in the
theory with running coupling in general is determined by
renormalon~\cite{t'Hooft,Parisi,David,Mueller85}.  The renewed
interest in this kind of asymptotic estimates have been demonstrated
in last few
years~\cite{West,Brow,VZre,Grun,Mueller,Beneke,Broa,Vainshtain,Maxwel}.
It results even in attempts \cite{BBB} to use renormalon for
calculation of experimentally measurable quantities. Finally, the
recent explosion of activity in the renormalon business has been
stimulated by the understanding that the former may be related with
the anomalously large ($\sim \Lambda_{QCD}/Q$) nonperturbative
corrections in hadronic shape 
variables~\cite{ES1,ES2,ES3,ZNPB,ES4,Zakharov}.

The accurate determination of renormalon-type asymptotics turns out to
be not so simple problem. It was recognised
\cite{Grun,Mueller,Beneke,Vainshtain,BeS} that the overall
normalisation of the renormalon could not be found without taking into
account of all terms of the expansion of, e.g., the Gell-Mann--Low
function. However, surprisingly up to now nobody have tried to sum up
this series of corrections to the renormalon.

Therefore the main aim of this paper will be to discuss the possible 
consequences of
summation of contributions from the arbitrary high order insertions to
the dressed gluon line.  As we demonstrate both by diagrammatic
consideration and by direct analytical calculation, each new type of
insertions generates the correction to renormalon of the order of
$\sim 1$. However, the $k$-th correction to the asymptotics for large
$k$ is not expected to have any $k!$ enhancement. Thus at least the
series of corrections to the amplitude of renormalon asymptotics is
not the asymptotic series. On the other hand, the summation of 
this series leads
to the sufficient change of the high order behaviour of the usual
perturbative series. We show that for the UV renormalon taking into
account of the high order corrections may change the $N$-th term of
the perturbation theory by the factor $N^\delta$ (although with
$\delta$ probably unknown even for QED). The $m$-th correction to the 
IR renormalon diverges like $(-2)^m$. Due to this divergency the IR
renormalon does not exist in the usual sense. We propose however the
proper rearrangement of the series which allows to avoid this
difficulty. The IR renormalon now will be associated with certain
non-Borel-summable series but the coefficients of this series
themselves are the Borel-summable series in $\alpha_s$. Finally, the 
event shape renormalons, which are the most powerful among other 
renormalons, may be quite stable under taking
into account of the high order insertions. The corresponding $m$-th
correction is expected to decrease like $(-2)^{-m}$.

In accordance with their name renormalons are connected with the
running of the coupling constant. Therefore in considering of the high
order corrections to renormalon one is faced with the problem of
renormalization scheme independent definition of the coupling. This
problem may be solved quite naturally in QED. That is why, in
particular, all the diagrammatic examples of this paper will be of the
QED--type. The generalization to the QCD case is usually done in some
heuristic way like e.g. the so called 'naive nonabelization'.
Nevertheless, the first attempts to find the renormalization scheme
independent definition of the effective charge in QCD have been done
recently \cite{Watson}. Also we will not usually make much difference
between photon and gluon, thus making the tacit assumption that the
difference between theories is hidden somewhere in the coefficients
$b_0,b_1,b_2, \dots$ of the Gell-Mann-Low equation.

The contribution of the diagrams with exchange of one soft
gluon(photon) to some ``physical'' quantity which accounts for the
infrared(IR) renormalon has the generic form
\bq\label{renIR}
R_{IR} = \int_{k\ll Q} \alpha(k) \fr{ d^4k}{Q^4} \ \ .
\ee
The Feynman graphs corresponding to this quantity are shown in
fig.~1. 
\input epsf
\begin{figure}[ht]
\epsfysize=5 truecm
\epsfxsize=10 truecm
\centerline{\epsffile{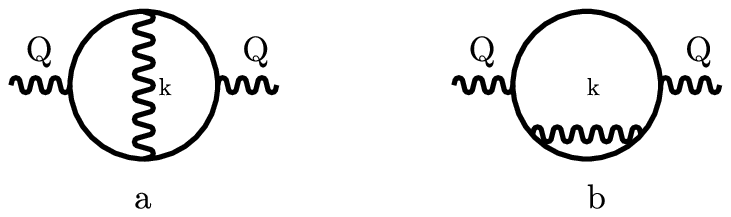}}
{\footnotesize {\bf Fig. 1} The renormalon graphs with exchange of
one gluon. The internal gluon line will be dressed in the
following figures.}
\end{figure}
Similarly the exchange of very hard gluon(photon) generates
the ultraviolet(UV) renormalon
\bq\label{renUV}
R_{UV} = \int_{k\gg Q} \alpha(k) \fr{Q^2 d^4k}{k^6} \ \ .
\ee
Now only the diagram of fig.~1a contribute. During the last two years
it was understood \cite{Vainshtain,BeS} that the traditional UV
renormalon (\ref{renUV}) with exchange of only one hard photon does
not give the largest contribution to the asymptotics. The diagrams
with exchange of at least two photons turns out to be much more
important. We will return to discussion of this new UV renormalon
later while now our consideration of the traditional UV renormalon is
of mainly methodological importance.

In (\ref{renIR},\ref{renUV}) we have written down the effective
running coupling constant $\alpha(k) = \alpha_{eff}(k)$, which
(at least for QED, but see also \cite{Watson}) is trivially connected
with the transverse part of the gluon propagator.  The function
$\alpha(k)$ satisfies the RG equation:
\begin{eqnarray}\label{GML}
\fr{d\alpha}{dx}=
b_0 \alpha^2 +b_1 \alpha^3 +b_2 \alpha^4 + ... \ \ , 
\ \ \ \ x= \ln\left( {Q^2}/{k^2} \right) \ \ \ .
\end{eqnarray}
It is to be noted here that we have
fixed the renormalization scheme by considering the effective
charge. Thus our coefficients $b_2,b_3,...$ are neither the free
parameters, nor the known, e.g., for $\overline{MS}$ scheme,
$b_2(\overline{MS}), \ b_3(\overline{MS})$ . 
At first stage one may neglect $b_1,b_2,\, ...$ in (\ref{GML})
\begin{eqnarray}\label{rendef}
R_{IR} = \fr{2}{\alpha_0} \int_0^{\infty} \alpha(x) e^{-2x}dx =
\int_0^{\infty} 
\fr{e^{-2x}}{1-b_0\alpha_0 x} 2 dx = 
\sum_{N=0} \left(
\fr{b_0\alpha_0}{2} \right)^N N! \ \ \ .
\end{eqnarray}
Here we have chosen some convenient overall normalization of the
renormalon.  We will consider now only the asymptotics of the
perturbation theory and leave the issue of the nonperturbative
ambiguity of the integral (\ref{rendef}) due to the Landau pole to the
very end of the paper. The integral (\ref{renIR}) describes adequately
the contribution of a certain chain of Feynman diagrams only for $k\ll
Q$ . It is seen from (\ref{rendef}) that the main contribution to the
$N$-th order of the expansion comes from $k^2 \sim Q^2 e^{-N/2}$. Thus
the renormalon contributions to the first few terms of perturbation
theory are completely irrelevant. On the other hand, for sufficiently
large $Q$ a lot of terms of the expansion (\ref{rendef}) come from the
region $\Lambda_{QCD}^2 \ll k^2 \ll Q^2$, where the effective charge is
small and the perturbative approach for calculation of $\alpha_{eff}$
(\ref{GML}) seems to be useful.

Very similar calculation for the UV renormalon (\ref{renUV}) leads to
the same result as (\ref{rendef}) up to trivial replacement
$\fr{b_0}{2} \rightarrow -b_0$. This means on the one hand that the UV
renormalon asymptotics in general is much more important (in $2^N$
times) than the IR one. On the other hand one may see that depending
on the sign of $b_0$ the series associated with one of the two
renormalons should be non-Borel summable.

{\footnotesize
\input epsf
\begin{figure}[ht]
\epsfysize=5 truecm
\epsfxsize=15 truecm
\centerline{\epsffile{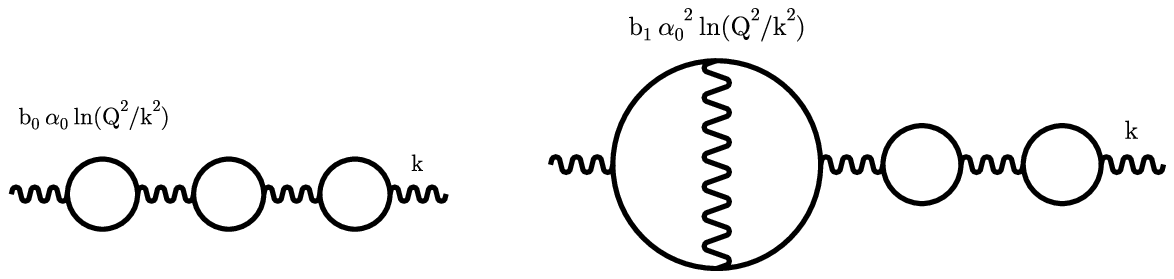}}
\begin{tabbing} 
{\bf Fig. 2} The simplest chain of diagrams $\ \ \ \  $
               \= {\bf Fig. 3} The example of diagram  with  \\
corresponding to the renormalization  
                          \>  two -- loop insertion into soft gluon line.\\
of the soft gluon line. Each bubble  \>                                 \\
generates the factor $b_0 \alpha_0\ln{(Q^2/k^2)}$. \>                  
\end{tabbing}    
\end{figure}
}

\section{Diagrammatic examples}\label{sec2}

The main part of the current interest in renormalons seems to be
concentrated on the consideration of their relation with
nonperturbative corrections to various observables. However, just this
part of the renormalon business suffers mostly on the hard to
establish assumption about the enhancement of nonperturbative
corrections compared to the typical uncertainty of renormalon 
resummation. The only free of phenomenological input use of renormalon
is for the approximate (with $1/N$ as a small expansion parameter,
where $N$ is the number of the term of the perturbative series)
calculation of the high order contribution of the usual perturbation
theory. Therefore, before passing to straightforward but rather formal
manipulations with the RG equation (\ref{GML}) let us illustrate the
role of complicated contributions to renormalon by the few explicit
estimates of Feynman graphs.

Moreover, the experience of this diagrammatic consideration will help
us in the following to distinguish, to what extent the factorial
growth of the series is connected with the true
infra-red(ultra-violet) physics or appears due to a simple
combinatorics.

In this section we will consider only the
IR renormalon. Generalization for the UV one is straightforward. The
fig.~2 shows the chain of diagrams corresponding to (\ref{rendef}).
We show only the QED--type diagrams without gluon
self-interaction. Each of the $N$ bubbles from fig.~2 generates the
factor $b_0\alpha_0 \ln\left( {Q^2}/{k^2} \right) $ in the integrand
of (\ref{renIR}),(\ref{rendef}).  The difference between QCD and QED
may be thought to be hidden in the factor $b_0$, accompanying the
single bubble.

Now let us replace two of the simple bubbles by the more
complicated diagram of fig.~3.  The two loop bubbles generate
the factor $b_1 \alpha_0^2 \ln\left( {Q^2}/{k^2} \right) $ in
the integrand, which has one power of large logarithm less (or
one $\alpha_0$ more) than the leading order contribution
(\ref{rendef}).  However, a large combinatorial factor $N$
appears due to a number of permutations of the second order
bubble among the simple bubbles, leading to
\bq\label{b1sum}
N b_1 \alpha_0^2 \ln\left( \fr{Q^2}{k^2} \right) \left[ b_0
\alpha_0 \ln\left( \fr{Q^2}{k^2} \right) \right]^{N-2}
\rightarrow \left( \fr{b_0\alpha_0}{2}\right)^N N!
\fr{2b_1}{b_0^2} \ .
\ee
Thus we see that taking into account one second order insertion
into the soft gluon line leads to the correction of the order of
one to the trivial asymptotics (\ref{rendef}).

Consider now the more complicated diagram of fig.~4 with
dressing of the internal  gluon line of the second order bubble.
To this end it is natural to write down explicitly the last
integration over internal momentum of the two loop diagram  
\bq\label{b1ll}
b_1 \alpha_0^2 \int_{k^2}^{Q^2} \left[ b_0
\alpha_0 \ln\left( \fr{Q^2}{q^2} \right) \right]^n
\fr{dq^2}{q^2} = \fr{1}{n+1} b_1 \alpha_0^2 \ln\left(
\fr{Q^2}{k^2} \right) \left[ b_0 
\alpha_0 \ln\left( \fr{Q^2}{k^2} \right) \right]^n \ .
\ee
Thus up to the overall factor $\fr{1}{n+1}$ the contribution of
diagram of fig.~4 coincides with 
\input epsf
\begin{figure}[ht]
\epsfysize=4.7 truecm
\centerline{\epsffile{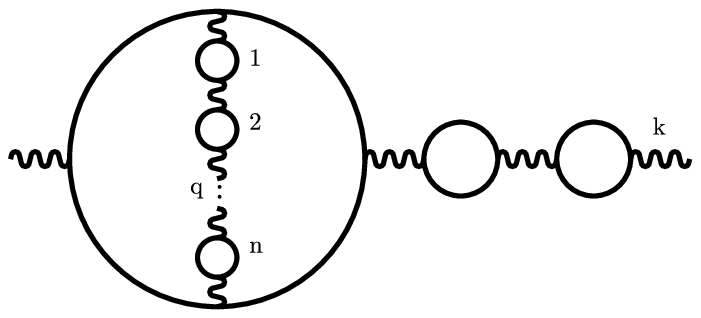}}
{\footnotesize {\bf Fig. 4} The dressing of internal gluon line of the
second order bubble by $n$ simple bubbles.}
\label{fig.4}
\end{figure}
\input epsf
\begin{figure}[ht]
\epsfysize=4.7 truecm
\centerline{\epsffile{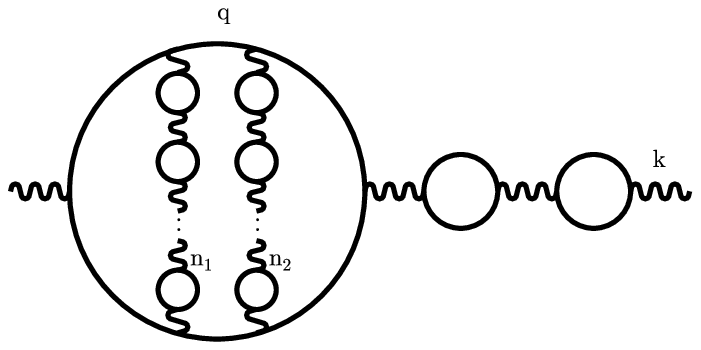}}
{\footnotesize {\bf Fig. 5} Three loop insertion with dressing of
two internal gluon lines by the simple chains of bubbles.  The
summation over $n_1$ and $n_2$ allows to compensate all extra
$\alpha$-s.}
\label{fig.5}
\end{figure}
that of fig.~3. Summation over
$n$ naturally leads to $\ln(N)$. Taking into account 
a number of large bubbles of fig.~4 allows to exponentiate the
correction
\bq
\label{b1lexp}
\left( \fr{b_0\alpha_0}{2}\right)^N N!
\exp \left( \fr{2b_1}{b_0^2} \ln(N) \right) = \left(
\fr{b_0\alpha_0}{2}\right)^N N^{\fr{2b_1}{b_0^2}} N! \ .
\ee
This is the generally recognised expression for the IR
renormalon. Our argumentation up to this stage repeats the line
of reasoning of the paper \cite{VZre}.  However, the argument of
the exponent in (\ref{b1lexp}) was found with the
$\sim~1/\ln(N)$ accuracy and therefore the nontrivial overall
factor as well as the function of $N$, weaker than $N^{\gamma}$,
may appear in (\ref{b1lexp}).

Now let us consider the three loop correction 
(fig.~5 with $n_1=n_2=0$). This contribution generates the
factor $b_2 \alpha_0^3 \ln\left( {Q^2}/{k^2} \right) \ $ in
the integrand of (\ref{rendef}).  Thus here we have two extra
$\alpha_0$ which at first glance could not be compensated by one
combinatorial $N$ and hence the diagram of fig.~5 seems
to generate only the $\sim 1/N$ correction to renormalon. 
However, let us see, what happens if one dresses the internal
gluon lines of the three loop diagram. Now summation over the
number of trivial insertions $n_1,n_2$ gives:
\begin{eqnarray}\label{b2nn}
\, b_2 \alpha_0 \ln\left( \fr{Q^2}{k^2} \right) 
\sum_{n_1,n_2}
\alpha_0^2 \fr{N-n_1-n_2-2}{n_1+n_2+1}  
\sim b_2 \alpha_0 \ln\left( \fr{Q^2}{k^2} \right) \times
(\alpha_0 N)^2 
\ . 
\end{eqnarray}
Here the factor $(n_1+n_2+1)^{-1}$ appears after integration
over the internal momentum of the large bubble, while
$(N-n_1-n_2-2)$ accounts for the combinatorics.  We see that
after dressing of all gluon lines the three loop ($\sim b_2$)
diagram generates the correction to renormalon of the order of
$\sim 1$. One can easily show that four loop ($\sim b_3$), five
loop ($\sim b_4$) etc. diagrams generate the corrections of the
same order of magnitude.  Previously the analogous proof of the
importance of the high loop corrections was done by
Mueller\cite{private} but this result was not published.

\section{Upon summation of the corrections for the UV 
renormalon}\label{sec3}

It is easy to integrate formally the renormalization group
equation (\ref{GML}) \bq\label{GMLi} -\fr{1}{\alpha}+\fr{1}{\alpha_0}-
\fr{b_1}{b_0}\ln\left( \fr{\alpha}{\alpha_0} \right) - c_2
(\alpha-\alpha_0)-c_3 (\alpha^2-\alpha_0^2)-...  =b_0 x \ , \ee where,
$c_2=b_2/b_0-b_1^2/b_0^2 \ , \ c_3=(b_3/b_0-2b_2
b_1/b_0^2+b_1^3/b_0^3)/2, \ ...$ . For arbitrary $k$ one has
$c_k=(b_k/b_0 - ...)/(k-1)$.  This ``exact'' solution still is too
informative for us. First of all, most of the terms containing
$\alpha_0$ in (\ref{GMLi}), namely $c_2\alpha_0, c_3\alpha_0^2,
c_4\alpha_0^3, ...$ will contribute only to the $\sim 1/N$ corrections
to the asymptotics. Therefore we may write
\begin{eqnarray}\label{renalpha}
\alpha= \fr{\alpha_0}{1-b_0 \alpha_0 x- ({b_1}/{b_0})
\alpha_0\ln (
{\displaystyle {\alpha}/{\alpha_0}} ) - c_2\alpha \alpha_0
-c_3\alpha^2\alpha_0 - ...} \ \ .
\end{eqnarray}
Let us introduce now the truncated running coupling
\bq\label{renalt}
{\alpha_t} = \fr{\alpha_0}{1-b_0
\alpha_0 x- ({b_1}/{b_0})\alpha_0\ln (
{\displaystyle {\alpha_t}/{\alpha_0}} ) } \ \ .
\ee
Now one may again formally expand the exact $\alpha$ (\ref{renalpha})
in the series
\bq\label{alphal}
\alpha = \alpha_t \bigg( 1+\beta_2\alpha_t^2 +
\beta_3\alpha_t^3 + \, ...\, \bigg) \ ,
\ee
The explicit formulas for a first few coefficients $\beta_k$ may be
easily found. However we will be interested only in the large-$k$
asymptotic behaviour of $\beta_k$. Naturally for our definition of the
effective charge (\ref{GML}) the coefficients $b_k$ themselves form
the asymptotic series $b_k \sim k!$. Due to that one easily finds the
asymptotics of $\beta_k$ in (\ref{alphal}).
\bq\label{beta}
\beta_k = \fr{b_k}{k b_0}\left( 1+O\bigg(\fr{1}{k}\bigg) \right) 
\ \ \ .
\ee

In the previous section we have shown diagrammatically that
contributions to the asymptotics from the high order terms of the
RG equation (\ref{GML}) $\sim b_2,b_3,b_4, ...$ (or now
$\sim \beta_2,\beta_3,\beta_4, ...$ (\ref{alphal})) are not
small. However, the contribution induced by the second (two loop) term
$b_1\alpha^3$ play an outstanding role due to the additional
enhancement by $\ln (N)$. This is the reason for taking the truncated
coupling $\alpha_t$ (\ref{renalt}) as the expansion parameter in
(\ref{alphal}). 

The expression (\ref{renalt}) itself is the transcendental equation
for function $\alpha_t(\alpha_0)$ which may be solved iteratively. In
ref. \cite{FS} the IR renormalon asymptotics with $\alpha_{eff}$ in
(\ref{renIR}) replaced by the only truncated $\alpha_t$ (\ref{renalt})
has been considered in details. This calculation turns out to be
surprisingly sophisticated. For any finite number of iterations in the
transcendental equation (\ref{renalt}) the asymptotics contains some
new functional dependence on $N$ compared to the usual IR renormalon
(\ref{b1lexp}). The generally accepted result is restored only after
performing the infinite number of iterations in
(\ref{renalt}). However, we will skip now the discussion of the role 
of the first few terms of the series (\ref{alphal}) and concentrate our
attention on $\beta_k$ with very large $k$. In this case it will be
enough to make only one iteration in (\ref{renalt}) (namely
$\ln(\alpha_t/\alpha_0) = - \ln(1-b_0 \alpha_0 x)$ in (\ref{renalt})).

Consider first the UV renormalon. Moreover, let us restrict ourselves
on the UV renormalon for pure QED. The first important observation
(see \cite{Vainshtain,FS}) is that the series of corrections to
renormalon generated by (\ref{alphal}) although have no any reasonable
small parameter but also is not the asymptotic series. In order to
show that it is so consider for the moment the simplified version of
(\ref{renalt},\ref{alphal}) with $b_1=0$ . Now one has immediately for
the UV renormalon (\ref{renUV})
\bq\label{demo}
\sum_m \fr{1}{\alpha_0} \int_0^{\infty} \beta_m 
\left( \fr{\alpha_0}{1+b_0\alpha_0 y}\right)^{m+1} e^{-y} dy =
\sum_m \alpha^N (-b_0)^N N! \fr{\beta_m}{(-b_0)^m m!} \ .
\ee
We see that though $\beta_m$ themselves are determined by the UV
renormalon and therefore have the form $\beta_m \sim m^{\gamma}
(-b_0)^m m!$ the two most dangerous factors from the $\beta_m$, the
$m!$ and $(-b_0)^m$, have been removed from the series.

Up to now we have chosen rather arbitrarily the overall normalization
of renormalon. In order to go further in understanding the role of
high order corrections to UV renormalon chain we have to specify the
normalization. Consider following ref.~\cite{Vainshtain} (and almost
everybody others in the renormalon business) the correlation function
of two electromagnetic currents (see also the eqs. (1,2) of
ref. \cite{Vainshtain} connecting this quantity with $R_{e^+e^-
\rightarrow hadrons}$)
\bq\label{emcorr}
\Pi_{\mu\nu} =i \int dx e^{iqx}\langle 0| T\{ j_\mu(x) j_\nu(0) 
\}|0 \rangle = (q_\mu q_\nu - g_{\mu\nu} q^2) \Pi(Q^2) \ \ .
\ee
We will be naturally interested in $\Pi(Q^2)$ in the Euclidean domain.

The simple calculation of the diagram fig.~1a for $k\gg Q$ gives
\bq\label{Pidef}
\Pi(Q^2) = \fr{N_f}{12\pi^2} \left\{
\ln\left(\fr{\mu^2}{Q^2}\right) + ... - 
\fr{1}{3\pi} \int \alpha (k) \ln\left(\fr{k^2}{Q^2}\right)
\fr{Q^2dk^2}{k^4} \right\} \ \ .
\ee 
Here the first term in brackets is the parton model prediction, while
the integral is expected to generate the asymptotic series of the
perturbation theory. One may substitute into (\ref{Pidef}) the one
loop running coupling constant and easily reproduce the
``traditional'' (before \cite{Vainshtain,BeS}) UV renormalon
\cite{BeBe}. By dots in (\ref{Pidef}) we have denoted the rest part of
the perturbative series which is not included into the leading UV
renormalon (and is expected to be much smaller than the UV
renormalon).

The polarization operator $\Pi$ is trivially connected with the
$\beta$-function for the effective charge (\ref{GML})
\bq\label{x7}
4\pi \left( Q^2\fr{d\Pi}{dQ^2}\right)_{\mu^2=Q^2} =
\sum b_n [\alpha(Q)]^n \ \ \ .
\ee
However, because we are interested only in the
asymptotics we may find directly from the equation
(\ref{GML}) that
\bq\label{x8}
\fr{d}{d\ln(Q^2)}  n!(-b_0\alpha)^n 
= (n+1)! (-b_0\alpha)^{n+1} \left( 1+ 
O\bigg( \fr{1}{n}\bigg)\right) \ \ .
\ee
Thus for our purposes
\bq\label{Piser}
4\pi \Pi(Q^2) = \sum b_n [\alpha(Q)]^n
\ee 
Now we have to substitute the series for the effective charge
$\alpha(k)$ (\ref{alphal}) into the integral (\ref{Pidef}) and expand
the result in series in $\alpha(Q)$. The effective method which allows
to find the coefficients of such expansion was developed in
\cite{FS}. First of all it is convenient to introduce the new
variables
\bq\label{newvar}
a=-b_0 \alpha_0 \ \ \ , \ \ \ \beta = 
-{b_1}/{b_0^2} \ \ \ .
\ee
For QED $b_0=-\fr{N_f}{3\pi}\ , \ b_1=-\fr{N_f}{4\pi^2}$ and
$\beta=\fr{9}{4N_f}$. With this new variables the truncated effective
charge in the first nontrivial approximation takes the form
\bq\label{at}
\alpha_t = -\fr{1}{b_0} \ \fr{a}{1-ax+\beta a \ln(1-ax)}
\ \ \ ; \ \ x=\ln\left( \fr{k^2}{Q^2}\right) \ \ .
\ee
The contribution of the $n$-th term of the formal expansion
(\ref{alphal}) to the polarization operator (\ref{Pidef}) is of the
form
\bq\label{Pin}
\Pi_n= \fr{-\beta_n}{12\pi^2 (-b_0)^n}
\int_0^\infty \left[ \fr{a}{1-ax+\beta a \ln(1-ax)} \right]^{n+1} 
x e^{-x} dx \ \ \ .
\ee
Note that this is the $n$-th term of the expansion in the series in
truncated running coupling $\alpha_t(k)$ and it still contains the
whole series in $\alpha(Q)$. As before we are looking for the $N$-th
term of the series in $\alpha(Q)$. Below we will see that the most
important contribution comes from $n \sim N$. The simple binomial
expansion now gives
\begin{eqnarray}\label{bin}
\Pi_n = \fr{-\beta_n}{12\pi^2 (-b_0)^n}
\int \left[ \fr{a}{1-ax} \right]^{n+1}
\sum_{m=0} \fr{(n+m)!}{m!n!} \left[ \fr{-\beta a
\ln(1-ax)}{1-ax} 
\right]^m x e^{-x} dx =  \nonumber \\
= \fr{-\beta_n}{12\pi^2 (-b_0)^n}
\int 
 \fr{a^n}{(1-ax)^{n+1}}
\sum_{m=0} \fr{(n+m)!}{m!n!} 
\left[ \fr{-\beta a\ln(1-ax)}{1-ax} 
\right]^m e^{-x} dx  . 
\end{eqnarray}
Here in the second equality we have used that
\bq\label{trtrtr}
\fr{ax}{1-ax}=\fr{1}{1-ax}-1
\ee
and than have neglected the $-1$ (one may easily see 
that this less singular contribution will lead to the $\sim
1/N$ correction to the asymptotics).

For estimation of the $N$-th order contribution of perturbation theory
we will use the formula for $N$-th term of the expansion of the
integral in powers of $a$
\bq\label{formula} 
\left\{ \int \fr{e^{-t}dt}{1-at}
\left[ \fr{\beta a}{1-at} \right]^k \left[ \ln \fr{1}{1-at} \right]^m
\right\}_{N} = a^N N! \fr{\beta^k}{k!} \left[ \ln \fr{N}{k} \right]^m
 \ .  
\ee
Here both $m$ and $k$ are supposed to be large.
In order to derive (\ref{formula}) one has to use
the asymptotics of gamma-function together with the trivial identity
\bq\label{vareps} 
\big( \ln(p) \big)^n = \lim_{\varepsilon \rightarrow
0} \left( \fr{\partial}{\partial\varepsilon} \right)^n p^{\varepsilon}
\ .  
\ee

Formula (\ref{formula}) allows now to calculate the $N$-th term of the
expansion of (\ref{bin}) in series in $\alpha_0^N$ ($a^N$, $a=-b_0 
\alpha_0$)
\begin{eqnarray}\label{PinN}
\big\{ \Pi_n\big\}_N &=& \fr{-\beta_n}{12\pi^2 (-b_0)^n}
a^N N! \sum_{m=0} \fr{\beta^m}{(n+m)!}
\fr{(n+m)!}{m!n!} \left[ \ln\fr{N}{n+m}\right]^m=
 \nonumber \\
&=& - \fr{1}{12\pi^2} (-b_0\alpha_0)^N \fr{\beta_n}{(-b_0)^n n!}
\left( \fr{N}{n}\right)^{\beta} \ \ .
\end{eqnarray}
Here we have neglected $m$ compared to $n$ in the argument of
logarithm because effectively $m \sim \ln(N)$ while as we will see in
the moment $n\sim N$. Combining together (\ref{Piser},\ref{PinN})
and the asymptotics of $\beta_k$ (\ref{beta}) one finds the 
equation for $b_N$
\bq\label{bNeq}
b_N= -\fr{1}{3\pi b_0} (-b_0)^N N^\beta N! 
\sum_{n<N} \fr{1}{n} \ \fr{b_n}{(-b_0)^n n^\beta n!} \ \ .
\ee
Here $\beta = -b_1/b_0^2 = 9/(4N_f)$ for QED. This equation is even
further simplified after substitution
\begin{eqnarray}\label{cNeq}
b_N &=& (-b_0)^N N^\beta N! c_N \ \ , \\
c_N &=& \sum_{n<N} \fr{c_n}{n(-3\pi b_0)} = 
\fr{1}{N_f} \sum_{n<N} \fr{c_n}{n} \left( 1+ O\left(\fr{1}{n}
\right) \right) \ \ \ . \nonumber
\end{eqnarray}
The solution to this last equation is evident $c_N = const\times 
N^{1/N_f}$
and for $b_N$ one has
\bq\label{bNfin}
b_N = const (-b_0)^N N^{\fr{9}{4N_f}+\fr{1}{N_f}} N! \ \ .
\ee
So we arrived at the surprising result: {\it taking into account all
possible insertions to the renormalon chain allowed to change the
power of $N$ in the asymptotics, which for many years was thought to
be determined by only the two terms of the Gell-Mann--Low
$\beta$-function $b_0$ and $b_1$ .}

The overall constant in (\ref{bNfin}) could not be found in closed
form. In terms of the equation (\ref{cNeq}) this constant is
determined by the initial condition at $N\sim 1$. But for $N\sim 1$
the equation (\ref{cNeq}) is no more valid as we have indicated
explicitly by writing $(1+O(1/n))$.

This reduction to the initial value problem shows how one may
reformulate the problem of the overall normalization of
renormalon. Finally, the solution looks almost like tautology. One is
allowed to look for the renormalon asymptotics in the form
\bq\label{Rfin}
R_N = A_n N^\delta (-b_0)^N N! \ \ \ .
\ee
Here $n$ is the number of terms of the perturbation theory which were
calculated {\it explicitly} ($n<N$). The normalization constant $A_n$
may be found with only the $\sim 1/n$ accuracy. This is the important
difference between the renormalon and instanton \cite{Lipatov}
induced asymptotics. For instantons not only the overall amplitude of
the asymptotics is known but also the $\sim 1/N$ corrections to this
asymptotics were considered \cite{FS1,FS2} (see also \cite{Balitsky}).

However, one may consider the renormalon calculus only as a way to
extend the explicit perturbative calculations by one more approximate
term (and this is the most straightforward application of
renormalon). In this approach the equation (\ref{Rfin}) still is quite
informative. It shows that after the explicit calculation of $N$ terms
of perturbation theory one will immediately found the $N+1$-st term
with at least the $\sim 1/N^2$ accuracy.

\section{UV renormalons with many hard photons}\label{sec4}

Our formula (\ref{bNfin}) for asymptotics of the
coefficients $b_N$ would be a nice new result if published $3$-years
ago. However, as we have told in the introduction Vainshtein and
Zakharov in their preprint of April 1994 \cite{Vainshtain} have shown
that the traditional UV renormalon (\ref{renUV}) (fig. 1a) does not
give the largest contribution to the asymptotics. They have found a
series of new diagrams (starting from two three--loop diagrams) which
generate the asymptotics much larger than (\ref{renUV}). The authors
of \cite{Vainshtain} have used the OPE in order to find the
contribution to polarization operator of these new diagrams (see
Parisi \cite{Parisi}, who first proposed to use the OPE for renormalon
calculus). Finally, the new result for $\Pi(Q^2)$ reads
\bq\label{PiVZ}
\Pi_{UV}(Q^2) =
const \ \int_{k>Q} \left( \fr{\alpha(k)}{\alpha(Q)}
\right)^{2+\gamma} \fr{Q^2d^4k}{k^6}
\ee
with $\gamma$ for the QED case 
\bq\label{gamma}
\gamma = \fr{3}{N_f} \left( 
\sqrt{\left(\fr{2N_f+1}{6}\right)^2 +\fr{11}{4}}
-\fr{2N_f+1}{6}\right) \ \ \ .
\ee
Now it is easy to substitute the two loop running coupling
(\ref{renalt},\ref{at}) into (\ref{PiVZ}) and find the UV renormalon
of the form (\ref{Rfin}) with
\bq\label{delta}
\delta = 1+\gamma +\fr{9}{4N_f} \ \ \ .
\ee
However, it is clear that this result for $\delta$ will be changed
immediately if one substitutes the all--loop $\alpha_{eff}$ into
(\ref{PiVZ}) and repeats the simple calculation which we have
performed in the previous section for $\Pi(Q^2) $(\ref{Pidef}). The
magnitude of the correction to $\delta$ (\ref{delta}) now will depend
on the explicit value of the overall normalization of (\ref{PiVZ})
which may be extracted from \cite{Vainshtain,BeS}.

On the other hand, if we are going to substitute the exact running
coupling into (\ref{PiVZ}) we should also take into account all
possible corrections to this formula
\bq\label{PiVZeff}
\Pi_{UV}(Q^2) =
const \ \int_{k>Q} \left( \fr{\alpha_{eff}(k)}{\alpha(Q)}
\right)^{2+\gamma}
\big( 1+ \sum \pi_m \alpha_{eff}^m(k)\big) 
 \fr{Q^2d^4k}{k^6}
\ee
If the coefficients here behave like
\bq\label{pim}
\pi_m \sim \fr{b_m}{m}
\ee
for large $m$, taking into account of this series will change again
the exponent $\delta$ (\ref{delta},\ref{Rfin}). Moreover, if $\pi_m$
grows faster with $m$ than (\ref{pim}) it will be a catastrophe for
asymptotics as may be seen from (\ref{cNeq}). The eq. (\ref{pim}) is
likely the upper bound for the coefficients of the expansion
(\ref{PiVZeff}).

Unfortunately we do not see now the clear way to estimate the
asymptotics of the coefficients $\pi_m$ in (\ref{PiVZeff}). It is
rather probable that the corrections to $\delta$ due to the $\pi_m$ 
and $\alpha_{eff}(k)^{2+\gamma}$ will
compensate each other and the result of
refs. \cite{Vainshtain,BeS} will be restored. Furthermore, the authors
of \cite{Vainshtain} have demonstrated in their conclusions that they
are ready to meet any surprise from the high order corrections to the
renormalon. Therefore, even if such compensation do not take place and
$\delta$ (\ref{delta}) is changed this result will not be in complete
disagreement with \cite{Vainshtain}.

Anyway, after one finds the new $\delta$ from (\ref{PiVZeff}) the
overall normalization of the UV renormalon will be found with only
$\sim 1/n$ accuracy and only after the explicit calculation of the
first $n$ terms of the series. We write here $\sim 1/n$ although it
may be some power of it $\sim (1/n)^k$. Moreover, may be just the
better understanding of the high order contributions to renormalon
like in (\ref{PiVZeff}) may help to estimate, with what accuracy one
may found the $n+1$-st term of perturbation theory after explicit
calculation of $n$ terms?

\section{High 
order corrections for the IR renormalon and the best accuracy of
perturbative QCD}\label{sec5}

The IR renormalon have attracted much more interest during the last
few years than the UV one. This conclusion may be drawn even by simple
counting of the number of publications. It seems that most of the 
authors do not consider the UV renormalon as renormalon at all.

The reason for such asymmetry is quite evident. In QCD the UV
renormalon is Borel summable while the IR renormalon is
not. Physically this non--Borel--summability means that depending on
the details of chosen resummation procedure the obtained predictions
for observables will vary by some power corrections. Namely for the
case of light fermions and inclusive cross sections this correction
will be of the form $\sim (\Lambda_{QCD}/Q)^4$. There is absolutely no
way to find the amplitude of this nonperturbative $\sim Q^{-4}$
contribution from the perturbative renormalon calculus.

If there is no rigourous way to calculate the $\sim
(\Lambda_{QCD}/Q)^4$ correction one may try at least to extract them
from the comparison of theoretical prediction with the
experiment. Naively one may take say the experimental value of
$R_{e^+e^-\rightarrow hadrons}$ subtract the parton-model contribution
and $2-3$ known $\sim \alpha_s$ corrections and look for the power
corrections. However, this procedure certainly will not work at least
for sufficiently large $Q$\footnote{the success of the QCD sum rules
approach at low ($\sim 1GeV$) energies is usually considered as the
indication that the actual power corrections in QCD have some
additional enhancement compared to the typical uncertainty of
renormalon resummation. However, this enhancement will became less and
less important at high energies which only may be a subject of
perturbative QCD.}. The more or less reasonable procedure is the
following: one has to calculate a huge number
\bq\label{huge}
N_{IR}(Q) = \fr{2}{b_0\alpha(Q)}
\ee 
terms of perturbation theory ($N_{IR}$ is the function of $Q$ !) and
subtract them from the experimental ratio $R(Q)$. The rest will be the
needed $ (\Lambda /Q)^4$ correction.

The first problem (if not worry about the terrible analytical
calculations!) which will encounter us upon performing this procedure
is the UV renormalon. The series for UV renormalon blows up at
$N=1/(b_0\alpha(Q))$ -- twice before the critical value
(\ref{huge}). However the way to avoid this problem seems to be rather
clear. As we have seen (\ref{renIR},\ref{renUV}) the UV and IR
renormalons originate from the very different regions of variation of
the internal momentum in the diagram ($k$ in the fig.~1). Naturally
one may divide the integral into two parts $k<Q$ and $k>Q$ and than
obtain the result in the form
\bq\label{PiUVIR}
\Pi = \sum_{UV} (-b_0\alpha)^N N! \ +
\sum_{IR,N<N_{IR}} \left(\fr{b_0}{2}\alpha\right)^N N! \ \ .
\ee
Here the first series is much larger than the second one but allows
the explicit(Borel) summation. So one has to sum up exactly the series
for the UV renormalon in (\ref{PiUVIR}) and then subtract this
resummed contribution from the experimental value of $R_{e^+e^-\rightarrow
hadrons}$.

Our goal is to reach the best accuracy of the perturbative QCD $\sim
(\Lambda/Q)^4$. Therefore the summation of the UV renormalon in
(\ref{PiUVIR}) also should be done with $\sim (\Lambda/Q)^4$
accuracy. This is also not so easy to do because even the smallest
term of the series for the UV renormalon is of the order of $\sim
(\Lambda/Q)^2$ -- much larger than the accuracy we want. For example,
if one simply substitutes the $1$-loop $\alpha(k)$ into naive
renormalon (\ref{Pidef}) the accuracy of the resummed value will be
only $O(\alpha^2)$ and so on. It is quite probable that in order 
to sum up
the UV renormalon with the $(\Lambda/Q)^4$ accuracy one still has to
calculate exactly about $N_{IR}=2/(b_0\alpha)$ terms of the series.

However, suppose that accurate enough summation of the UV renormalon
in (\ref{PiUVIR}) has been done. Let us see, what can we say about
the IR-renormalon part of the polarization operator in view of our
experience of working with high order corrections to renormalon chain?
The corresponding contribution to $\Pi$ (\ref{emcorr}-\ref{Piser}) now
has the form of the integral (\ref{renIR}). Simply repeating the
calculations (\ref{Pin}-\ref{bNeq}) one gets
\bq\label{PiIRN}
\big\{ \Pi_{IR}\big\}_N = const \left( \fr{b_0}{2}
\alpha\right)^N N^{-1+{2b_1}/{b_0^2}} N!
\sum_n \fr{\beta_n}{(b_0/2)^n n^{2b_1/b_0^2} n!} \ \ \ .
\ee
But $\beta_n\sim b_n$ now is determined by the first UV renormalon and
thus $\beta_n\sim (-b_0)^n n!$ . Therefore we see that the sum of the
multiloop corrections to the IR renormalon leads to the correction of
the relative order of magnitude
\bq\label{diver}
\sim \sum_n \fr{\beta_n}{(b_0/2)^n n!}
\sim \sum (-2)^n = \infty \ \ \ .
\ee
The series of the corrections to renormalon is not asymptotic, but is
ugly divergent. It looks like some interference of the IR
and UV renormalons.

Nevertheless, as we will see now the divergency itself of the series
(\ref{diver}) is not connected with the infrared physics. As we saw in
(\ref{rendef}) the $N!$ in the leading contribution to the IR
renormalon appears simply due to the $N$-th power of the large
logarithm $x^N = \ln (Q^2/k^2)^N$ . Therefore as we have told after
the eq. (\ref{rendef}) this trivial renormalon contribution falls down
into the infrared momentum region like $k^2\sim Q^2 e^{-N/2}$. On the
other hand, as we have seen in the section 2 for high order
corrections to the renormalon chain the same $N!$ appears not only due
to the power of logarithm but also due to the more and more
complicated combinatorics. This change in the origin of the $N!$ would
not be taken into account if the series of corrections to renormalon
was convergent. However, for divergent series of the kind of
(\ref{diver}) one should naturally to reformulate the method of
calculation of IR renormalon in order to be able to control from what
distances each contribution came. To this end let us rewrite the
renormalon (\ref{renIR}) in terms of the series in the powers of $\ln
(Q^2/k^2)$
\begin{eqnarray}\label{Piln}
\Pi_{IR}(Q) \sim \int \fr{k^2dk^2}{Q^4} \alpha(k) &=&
\sum_n \int  \fr{k^2dk^2}{Q^4} \alpha(Q)^{n+1} \left[
b_0 \ln\left( \fr{Q^2}{k^2}\right) \right]^n F_n(\alpha(Q)) 
\nonumber \\ 
&=& \sum_n  \alpha(Q)^{n+1} F_n(\alpha(Q)) 
\left(\fr{b_0}{2}\right)^n n! \ \ .
\end{eqnarray}
Now all the high order corrections to renormalon chain are hidden in
$F_n(\alpha)$. The function $F_n(\alpha)$ itself may be expanded in
the asymptotic series. But it would be the Borel summable series.

In the previous two sections we have tried to show that there is a lot
of open questions concerning the UV renormalon. Therefore, now we will
illustrate only by the simple toy example how the functions
$F_n(\alpha)$ (\ref{Piln}) will be found after one solves all this
``ultraviolet problems''. Suppose the running coupling $\alpha(k)$ in
(\ref{Piln}) is described by the simplest Borel integral
\begin{eqnarray}\label{toy}
\alpha(k) = \int_0^\infty \fr{\tilde{\alpha}}{1+b_0 \tilde{\alpha}(x)y} 
e^{-y} dy = \tilde{\alpha}
\sum_{n=0}^\infty (-b_0\tilde{\alpha})^n n! \ \ \ , \\
\tilde{\alpha}(x)=\fr{\alpha_0}{1-b_0 \alpha_0 x} \ , \ 
x=\ln\left( \fr{Q^2}{k^2}\right) \ , 
\ y=\ln\left( \fr{k'^2}{k^2}\right) \ . \nonumber
\end{eqnarray}
Here $\tilde{\alpha}(x)$ is simply the one loop running coupling for
the momentum $k$.  The trivial expansion in the series over $x$ now
gives
\bq\label{Ftoy}
F_n = \int_0^{\infty} \fr{e^{-y}}{(1+b_0\alpha_0 y)^{n+1}}
dy
\ee
Of course this $F_n(\alpha_0)$ may be expanded in the asymptotic
series in $\alpha_0$. For very large number $n$ the integral in
(\ref{Ftoy}) is even further simplified. For example for the last
$F_n$-s which one is allowed to work with (with $n\approx N_{IR}$
(\ref{huge})) the expression (\ref{Ftoy}) reduces to almost trivial
result $F_{N_{IR}}=1/3$.

\section{Event shape renormalons and conclusions}\label{sec6}

The IR renormalons for the event shape variables have became the
popular topic of high energy physics in last couple of years (see e.g
\cite{Zakharov} and references therein). It was recognized
that among the experimental values,
which can not be directly related to the usual operator product
expansion in Euclidean domain, one may find the quantities for which
the soft gluon contribution is much less suppressed than in
(\ref{renIR}). In particular, the corresponding integral may even
contain only the first power of $k$ instead of the usual fourth 
power (\ref{renIR})
\bq\label{renES}
R_{ES} = \int_{k\ll Q} \alpha(k) \fr{ dk}{Q} \ \ .
\ee
It is clear, that such contribution will generate the very
large $\sim \Lambda/Q$ nonperturbative correction. Like for other
renormalons this correction currently can not be found within {\it ab
initio} analytic calculation. One may suppose however, (as we have
done in fact in (\ref{huge})) that the smallest term of the series
associated with (\ref{renES}) may be considered as the proper order of
magnitude estimate of this $\sim \Lambda/Q$ correction. Then in order
to be able to extract from the experiment the pure nonperturbative
part one should calculate explicitly the
\bq\label{hugeES}
N_{ES}(Q) = \fr{1}{2b_0\alpha(Q)} \ \ 
\ee
terms of perturbation theory. Although, formally this number
$N_{ES}(Q)$ still is parametrically very large $\sim 1/\alpha$ , 
it is $4$ times
smaller than the corresponding value for the usual IR renormalon
(\ref{huge}).

Moreover, it is usually assumed \cite{ZNPB} that the actual
$\sim \Lambda/Q$ corrections may be sufficiently enhanced compared to
the smallest term of the series. Suppose that this is just the case and
the nonperturbative $\Lambda/Q$ contribution is in $C$ times enhanced
compared to the smallest term with some $C\gg 1$. In this case one 
may easily modify the eq. (\ref{hugeES})
\bq\label{hugeESm}
N_{ES}(Q) = \fr{1}{2b_0\alpha(Q)} 
-\sqrt{\fr{\ln(C)}{b_0\alpha(Q)}} \ \ .
\ee 
While writing this formula we still have assumed that $b_0\alpha
\ln(C)\ll 1$. Naturally the difference $\Delta N_{ES}$ between 
(\ref{hugeESm}) and (\ref{hugeES}) became relatively less important
for smaller coupling $\Delta N_{ES}/N_{ES} \sim \sqrt{\alpha \ln(C)}$.

Now let us return to the main subject of our paper and consider the
high order corrections to the Event-Shape renormalon. To 
obtain the Event-Shape renormalon (\ref{renES}) one has to consider
some rather special characteristic of the jet distribution (thrust
for example in \cite{ZNPB}). Therefore, it is natural to suppose
that the high order corrections to the effective coupling in 
(\ref{renES}) are still gathered by the traditional renormalons, and
especially by the UV one. Under this assumption, repeating again
the calculation (\ref{Pin}-\ref{bNeq}) one gets for the $N$-th 
term of the expansion of (\ref{renES})
\bq\label{RNES}
\big\{ R_{ES}\big\}_N \sim (2b_0\alpha)^N N^{\lambda} N!
\left\{\sum_{m=1} r_m\right\} \ \ ,
\ee 
where the multiloop corrections $r_m$ are expected to behave like
\bq\label{rmES}
r_m \sim \fr{\beta_m}{(2b_0)^m m!} \sim(-1/2)^m \ \ .
\ee
Thus we see that the series of multiloop corrections to the 
Event-Shape renormalon will naturally be described by the simple 
summable geometrical progression. However, this stability of the 
Event-Shape 
renormalon also reflects the fact that the best accuracy which in 
principle may be achieved within the pure perturbative
calculation for (\ref{renES}) is much worse than for the usual 
IR renormalon ( $\sim \Lambda/Q$ instead of $\sim (\Lambda/Q)^4$
respectively~).

To conclude, the goal of this paper was mainly to outline 
the problems which should be solved upon taking into account the
multiloop corrections to the renormalon chain.
In particular, we have shown that these high order
corrections play very different role for UV, IR and Event-Shape
renormalons. The renormalon contribution to high order terms of
perturbation theory for Event Shape variables are most hugely
divergent compared to other renormalons. However, just the 
Event-Shapes
renormalons turns out to be most stable under taking into
account the multiloop corrections to renormalon chain.
On the opposite side, the traditional IR renormalon is much
weaker than other renormalons, but, as we have shown in Section 5,
it is most sensitive to the inclusion of complicated insertions 
into the chain (divergent geometrical progression). Nevertheless,
one is still able to associate with the IR renormalon some 
perturbation--theory--like series. Only the coefficients of 
this series should themselves be the sums of Borel summable
series in $\alpha_s$. The special importance of the traditional
IR renormalon in QCD is due to the fact that it gives the 
principal limitation on the accuracy of pure perturbative 
calculations. Finally, the resummation of high order corrections
to any type of renormalons is sufficiently based on the use of
UV renormalon. In Section 3 we have shown, how taking into 
account of the high order corrections to the single UV 
renormalon chain may be reduced to solution of a simple
self-consistent equation. A considerable progress have been
achieved last years in calculating the UV renormalon asymptotics
with many hard photon(gluon) chains \cite{Vainshtain,BeS,Kivel}.
Nevertheless, in Section 4 we speculate about the possibility to
change the asymptotics again by taking into account the
multi-loop insertions into this many -- chain renormalon.  

The most direct use of renormalons, as well as any
other asymptotic estimates, will be to predict the value of 
$n+1$-st term of perturbation theory after the explicit calculation
of $n$ terms. However, contrary to the case of Instantons 
\cite{Lipatov}, the overall amplitude of renormalon itself can
not be found without accurate multiloop calculation. Moreover,
even if the asymptotics of perturbation theory will be found, the
corrections to it may be also very important. For example in QCD
\cite{FS2} the corrections to the asymptotics are of the order
of $\sim N_c^2/N$ which makes practically useless the asymptotic
estimates in this case. Another application of renormalons may
be to use some kind of renormalon resummation in order to extract 
the pure nonperturbative--QCD contributions from the real 
high-energy experiment. However, in this case one may
hope either on the possible (occasional) huge enhancement of the
nonperturbative corrections, or to wait a few decades before the 
needed number of terms of perturbation theory 
(\ref{huge},\ref{hugeES},\ref{hugeESm}) will be calculated
explicitly.

\vspace{0.2cm}

This work was supported by the Russian Foundation for
Fundamental Research under Grant 95-02-04607a.  The work of S.F.
has been supported by the INTAS Grant 93-2492-ext within the
program of ICFPM of support for young scientists. P.S. thanks
the hospitality of the Instituut Lorentz, Leiden, where this paper
was completed.

\end{document}